# A Discrete Spherical Harmonics Method for Radiative Transfer Analysis in Inhomogeneous Polarized Planar Atmosphere

By

Romuald Tapimo

omipat@yahoo.fr

Hervé Thierry Tagne Kamdem[*]

herve.kamdem@univ-dschang.org

and

David Yemele

david.yemele@univ-dschang.org

Laboratory of Mechanics & Modeling of Physics Systems

Department of Physics/Faculty of Science

University of Dschang, Cameroon

P.O. Box 67 Dschang, Cameroon

---

[*]Corresponding author: Hervé Thierry Tagne Kamdem
Phone: Mobile (00237) 697353450
E-mail: ttagne@gmail.com, herve.kamdem@univ-dschang.org

**Abstract**

A discrete spherical harmonics method is developed for the radiative transfer problem in inhomogeneous polarized planar atmosphere illuminated at the top by a collimated sunlight while the bottom reflects the radiation. The method expands both the Stokes vector and the phase matrix in a finite series of generalized spherical functions and the resulting vector radiative transfer equation is expressed in a set of polar directions. Hence, the polarized characteristics of the radiance within the atmosphere at any polar direction and azimuthal angle can be determined without linearization and/or interpolations. The spatial dependent of the problem is solved using the spectral Chebyshev method. The emergent and transmitted radiative intensity and the degree of polarization are predicted for both Rayleigh and Mie scattering. The discrete spherical harmonics method predictions using 36 streams are found in good agreement with benchmark literature results. The maximum deviation between the proposed method and literature results and for polar directions $|\mu| \geq 0.1$ is less than 0.5% and 0.9% for the Rayleigh and Mie scattering, respectively. These deviations for directions close to zero are about 3% and 10 % for Rayleigh and Mie scattering, respectively.

**1. Introduction**

The multiple scattering of any radiation by the earth atmosphere in all directions at all frequency of the incident beam generally gives birth to a light with a different state of polarization to that of the incident beam (Cornet et al., 2010). The solar radiation in traversing the atmosphere is reflected by the atmospheric particles and also at the level of the ground. The reflected radiation and the outgoing long-wave radiation escape to the space. Information about the atmosphere is collected in measuring that emerged solar radiation (De Rooij, 1985). These measurements should be compared to the computation results for checking the assumptions made theoretically (Hovenier, 1969). Knowing the scattering properties of the atmospheric particles the reflected radiance is calculated for a better interpretation of the observations (De Rooij, 1985). The interest of the interpretation of the observations aims to determine the vertical structure and the physical properties of the atmosphere (De Rooij, 1985; Ackerman and Stokes, 2003). The measurement of the emergent and transmitted solar radiation from the atmosphere and its comparison to computation results are very important, for example, to verify approximations done on the size distribution of the atmospheric particles as functions of the radius. The study of the multiple scattering of polarized light is also important in the study of the energy balance of the earth (Wauben et al., 1993), in the calculation of the infrared of planetary atmospheres (Wauben et al., 1993), in the climate study and the weather prediction (Ackerman and Stokes, 2003), in the remote sensing (Ackerman and Stokes, 2003; Marshak and Davis, 2005; Barlakas et al., 2016) and in the atmospheric optics.

The above mentioned applications of the polarization indicate that the investigation of the polarization characteristics of the light propagating through the atmosphere remains an active task for the researchers. The transport of the polarized radiation in the atmosphere is described by the radiative transfer equations (RTE). This equation is a set of four coupled equations of Stokes parameters that can be put in a vector form called vector radiative transfer equation (VRTE). Although this vector equation has been widely studied in literature, accurate and simple techniques are still needed even for the simplest one-dimensional geometry (Spurr, 2008; Budak et al., 2017). Many methods have been presented in the literature for solving the vector equation of the propagation of the polarization light in the atmosphere including the Monte Carlo method (Cornet et al., 2010; Muñoz and Mills, 2015; Barlakas et al., 2016), ), the "Facile" or $F_N$ method (Garcia and Siewert, 1989), the spherical harmonics and variant (Benassi et al., 1985; Garcia and Siewert, 1986; Budak and Korkin, 2006) and the discrete ordinates method and variant (Schulz et al., 1999; Rozanov and Kokhanovsky, 2006; Garcia and Siewert, 2011; Korkin, 2013; Collin et al., 2014),. Each of these methods presents the advantages and the drawbacks. The Monte Carlo method, generally adopted for benchmark solutions, requires a large computer memory and a high computational time. The $F_N$ method, which is also generally used for benchmark solutions, converges slowly at the domain boundaries leading to a high approximate order for accurate results. The spherical harmonics method can be used for predictions at any polar and zenith angles within the atmosphere. However, this method requires very high order of angular approximations to converge. The discrete ordinates method is easy to implement for polarized radiative transfer. However, it requires interpolation and/or linearization of the Stokes vector at an arbitrary polar angle.

The present work presents a simple and efficient method for radiative transfer in an inhomogeneous polarized planar atmosphere illuminated at the top by a collimated sunlight while the bottom reflects the radiation. The method called the discrete spherical harmonics method (DSHM), combines principles of the spherical harmonics method and the discrete ordinates method. First, both the Stokes vector and the phase matrix are expanded in a finite series of the generalized spherical functions. Second, the resulting nonlinear coupled differential equations of spatial variable are expressed in a set of polar angular directions. These equations are then put in the matrix form and solved numerically using the spectral Chebyshev method (STM). The emergent/transmitted radiative intensity and the degree of polarization predictions at any polar direction for both Rayleigh and Mie scattering of homogeneous/inhomogeneous polarized planar atmosphere are predicted and compared with available benchmark literature results.

## 2. Radiative transfer equation

In a scattering, absorbing, and emitting plane-parallel atmosphere with polarization effects the transport of the energy is described by the vector radiative transfer equation (VRTE) which can be derived by applying the energy conservation to an element of volume to yield (Chandrasekhar, 1960; Wauben et al., 1993)

$$\mu \frac{d\mathbf{I}(\tau,\mu,\varphi)}{d\tau} + \mathbf{I}(\tau,\mu,\varphi) = \frac{\omega(\tau)}{4\pi} \int_0^{2\pi} \int_{-1}^{1} \mathbf{Z}(\mu,\varphi,\mu',\varphi') \mathbf{I}(\tau,\mu',\varphi') d\mu' d\varphi' + \mathbf{S}_0(\tau), \qquad (1)$$

where $\mathbf{I} = [I,Q,U,V]^T$ is the total polarized radiation or the total Stokes vector. Its components are Stokes parameters representing the total intensity $I(\tau,\mu,\varphi)$, $Q(\tau,\mu,\varphi)$ and $U(\tau,\mu,\varphi)$ characterizing the linearly polarized radiation, and $V(\tau,\mu,\varphi)$ describing the circularly polarized radiation. They are defined such that the meridian plane acts as the reference plane (Chandrasekhar, 1960). $\mu \in [-1,1]$ is the cosine of the polar angle measured from the positive $\tau$ axis (positive values of $\mu$ represent downward radiation), $\varphi \in [0,2\pi]$ is the azimuthal angle which informs one on how the properties of the atmosphere vary over a plane, $\tau \in [0,\tau_L]$ is the optical depth measured from zero at the top of the atmosphere (TOA) to a value $\tau_L$ at the bottom of the atmosphere (BOA). The thermal term $\mathbf{S}_0(\tau)$ results from the emission by the atmosphere and is important in the energy balance of the earth (Wauben et al., 1993). The degree of polarization of the scattering light is calculated at any point within the atmosphere and at any angular direction by (Stam et al., 2006)

$$p(\tau,\mu,\varphi) = \frac{\sqrt{Q(\tau,\mu,\varphi)^2 + U(\tau,\mu,\varphi)^2 + V(\tau,\mu,\varphi)^2}}{I(\tau,\mu,\varphi)}. \qquad (2)$$

The degree of polarization can be viewed as a sum of the degree of linear polarization (Bohren and Clothiaux, 2006),

$$p_L(\tau,\mu,\varphi) = \frac{\sqrt{Q(\tau,\mu,\varphi)^2 + U(\tau,\mu,\varphi)^2}}{I(\tau,\mu,\varphi)}, \qquad (3)$$

and the degree of circular polarization

$$p_c(\tau,\mu,\varphi) = \frac{V(\tau,\mu,\varphi)}{I(\tau,\mu,\varphi)}. \tag{4}$$

It follows from Eq. (4) that positive values of $V(\tau,\mu,\varphi)$ correspond to right-circular polarization and negative values to left-circular polarization.

Consider a finite atmosphere illuminated at every point at the top by a unidirectional sunlight and bounded at the bottom by a reflecting ground. The incident radiation is in the direction given by $(\mu_0,\varphi_0)$ and has an incident flux vector $\boldsymbol{I}_0$. The phase matrix, $\mathbf{Z}(\mu,\varphi,\mu',\varphi')$, is assumed to be the same at any point in the slab while the albedo of single scattering changes from point to point. The dependence of the albedo of single scattering, $\omega(\tau)$, with the optical depth makes the atmosphere an inhomogeneous medium. When the sunlight traverses the atmosphere, it excites particles of the atmosphere to radiate secondary waves that are superposed to give the total scattered wave. As consequence of the illumination, the total radiation in the atmosphere is composed of two parts: the collimated or the direct component (Benassi et al., 1985; Mishchenko, 1990)

$$\boldsymbol{I}_c(\tau,\mu,\varphi) = \pi \boldsymbol{I}_0 \delta(\mu-\mu_0)\delta(\varphi-\varphi_0)e^{-\tau/\mu} \tag{5}$$

and the diffuse part, $\boldsymbol{I}_d(\tau,\mu,\varphi)$, so that one can express the total polarized radiation as

$$\boldsymbol{I}(\tau,\mu,\varphi) = \boldsymbol{I}_d(\tau,\mu,\varphi) + \pi \boldsymbol{I}_0 \delta(\mu-\mu_0)\delta(\varphi-\varphi_0)e^{-\tau/\mu}. \tag{6}$$

Inserting Eq. (6) into Eq. (1) yields to the following equation of the diffuse component

$$\mu\frac{d\boldsymbol{I}(\tau,\mu,\varphi)}{d\tau} + \boldsymbol{I}(\tau,\mu,\varphi) = \frac{\omega(\tau)}{4\pi}\int_0^{2\pi}\int_{-1}^{1}\mathbf{Z}(\mu,\varphi,\mu',\varphi')\boldsymbol{I}(\tau,\mu',\varphi')d\mu'd\varphi' + \boldsymbol{S}(\tau,\mu,\varphi), \tag{7}$$

where

$$\boldsymbol{S}(\tau,\mu,\varphi) = \boldsymbol{S}_0(\tau) + \frac{\omega(\tau)}{4}\mathbf{Z}(\mu,\varphi,\mu_0,\varphi_0)\boldsymbol{I}_0 e^{-\frac{\tau}{\mu_0}}. \tag{8}$$

In Eq. (7) and (8) the subscript $d$ has been dropped down and $\boldsymbol{I}(\tau,\mu,\varphi)$ denotes the diffuse part of the Stokes vector. For $\mu > 0$, the solution of Eq. (7) is sought under the boundary conditions (Garcia and Siewert, 1986)

$$I(0, \mu, \varphi) = 0 \tag{9}$$

and

$$I(\tau_0, -\mu, \varphi) = \mu_0 \lambda_0 \boldsymbol{T} \boldsymbol{I}_0 e^{-\frac{\tau_0}{\mu_0}} + \frac{\lambda_0}{\pi} \boldsymbol{T} \int_0^{2\pi} \int_0^1 I(\tau_0, \mu', \varphi') \mu' d\mu' d\varphi', \tag{10}$$

where $\lambda_0$ is the coefficient for Lambertian reflection or the surface albedo and $\boldsymbol{T}$ the matrix of reflectance.

### 3. The scattering matrix

The phase matrix contains the scattering properties of the atmosphere and describes how the intensity and the state of polarization change when a radiation propagating in an incident direction $(\mu', \varphi')$, after scattering propagates in the direction $(\mu, \varphi)$. It is related to the scattering matrix $\boldsymbol{F}(\Theta)$ by (Hovenier, 1969; Sievert, 1982; Collin et al., 2014)

$$\boldsymbol{Z}(\mu, \varphi, \mu', \varphi') = \boldsymbol{L}(\pi - \sigma) \boldsymbol{F}(\Theta) \boldsymbol{L}(-\sigma), \tag{11}$$

where the scattering angle $\Theta$ is the angle between the incoming and the outgoing beam defined as (Korkin, 2013: Collin et al., 2014)

$$\cos\Theta = \mu\mu' + \sqrt{1-\mu^2}\sqrt{1-\mu'^2}\cos(\varphi - \varphi'). \tag{12}$$

The matrices $\boldsymbol{L}(\pi - \sigma)$ and $\boldsymbol{L}(-\sigma)$ are that needed to rotate the meridian plane before and after the scattering into a local scattering plane. Thus, scattering matrix relates the Stokes parameters of incident beam and scattered beam, defined with respect to their scattering plane. The matrix of rotation is defined in terms of the rotation angle, $\sigma$, by (Hovenier, 1969; Korkin, 2013)

$$\boldsymbol{L}(\pi - \sigma) = \boldsymbol{L}(-\sigma) = \begin{bmatrix} 1 & 0 & 0 & 0 \\ 0 & \cos 2\sigma & -\sin 2\sigma & 0 \\ 0 & \sin 2\sigma & \cos 2\sigma & 0 \\ 0 & 0 & 0 & 1 \end{bmatrix}. \tag{13}$$

For a distribution of particles that scatter independently the phase matrix depends on the radius $r$, the scattering angle and may be obtained by the Mie theory. In general, this matrix has 16 nonzero elements. These elements are expressed in terms of the amplitude of the scattered light and can be found in detail in (Mishchenko et al., 2004). For some assumptions about the distribution of particles, it is possible to reduce the number of nonzero elements of the matrix. Assuming that in the atmosphere particles are of the same kind, are randomly oriented and have at least one plane of symmetry or particles and their mirror plane in the same number (Van de Hulst, 1981) this matrix takes the form (Van de Hulst, 1981; Mishchenko, 1990)

$$\boldsymbol{F}(r,\Theta) = \begin{bmatrix} F_1(r,\Theta) & F_2(r,\Theta) & 0 & 0 \\ F_2(r,\Theta) & F_5(r,\Theta) & 0 & 0 \\ 0 & 0 & F_3(r,\Theta) & F_4(r,\Theta) \\ 0 & 0 & -F_4(r,\Theta) & F_6(r,\Theta) \end{bmatrix}. \tag{14}$$

The calculation of these matrix elements are found in (Hansen and Travis, 1974; van de Hulst, 1981). In the independent scattering the transformation matrix is integrated over all particles in the distribution. More generally, in the radiative transfer analysis, it is preferable to use the normalized scattering matrix than the non-normalized one. The normalization constant is determined so that (Muñoz and Mills, 2015)

$$\frac{1}{4\pi} \int_{4\pi} F_1 d\Omega = 1.$$

Denoted here as **P**, the dimensionless elements of the normalized scattering matrix for a unit volume are calculated as (Hansen and Travis, 1974)

$$P_\alpha(\Theta) = \frac{\lambda^2}{\pi k_{sca}} \int_{r_1}^{r_2} F_\alpha(r,\Theta) n(r) dr = \frac{\lambda^3}{2\pi^2 k_{sca}} \int_{x_1}^{x_2} F_\alpha(x,\Theta) n(x) dx, \tag{15}$$

where $\alpha = 1,2,\dots 6$, $\lambda$ the wavelength of the illumination, $r$ the radius of the particle, $n(r)$ the particle size distribution per unit volume, $r_1$ and $r_2$ the smallest and the largest size of the distribution, $x = 2\pi r/\lambda$ the size parameter of the particle and $k_{sca}$ the scattering coefficient defined as (Dombrovsky and Baillis, 2010)

$$k_{sca} = \frac{\lambda^3}{8\pi^2} \int_{x_1}^{x_2} x^2 Q_{sca}(x) n(x) dx \tag{16}$$

For the Mie scattering and in the case of isotropic and homogeneous spherical or polydispersed spherical particles, $F_1 = F_5$, $F_3 = F_6$ (Evans and Stephens, 1991); the scattering matrix does not depend on the azimuthal angle (Van de Hulst, 1981). For the molecular scattering and the scattering by particles small compared to the wavelength of the illumination, Eq. (15) reads (Hovenier, 1971; Mischenco et al., 2006)

$$\boldsymbol{P}(\Theta) = \frac{3}{4}\begin{bmatrix} 1+\cos^2\Theta & -1+\cos^2\Theta & 0 & 0 \\ -1+\cos^2\Theta & 1+\cos^2\Theta & 0 & 0 \\ 0 & 0 & 2\cos\Theta & 0 \\ 0 & 0 & 0 & 2\cos\Theta \end{bmatrix}. \tag{17}$$

The phase matrix defined by Eq. (11) is expanded in a Fourier series as (Garcia and Siewert, 1986)

$$\boldsymbol{Z}(\mu,\mu',\varphi-\varphi') = \sum_{m=0}^{\infty} \boldsymbol{\Phi}_1^m(\varphi-\varphi')\boldsymbol{W}^m(\mu,\mu')\boldsymbol{D}_1 + \boldsymbol{\Phi}_2^m(\varphi-\varphi')\boldsymbol{W}^m(\mu,\mu')\boldsymbol{D}_2 \tag{18}$$

where

$$\boldsymbol{\Phi}_1^m(\xi) = (2-\delta_{0,m})diag\{\cos\xi,\cos\xi,\sin\xi,\sin\xi\}, \tag{19}$$

$$\boldsymbol{\Phi}_2^m(\xi) = (2-\delta_{0,m})diag\{-\sin\xi,-\sin\xi,\cos\xi,\cos\xi\}, \tag{20}$$

$$\boldsymbol{D}_1 = diag\{1,1,0,0\} \text{ and } \boldsymbol{D}_2 = diag\{0,0,1,1\} \tag{21}$$

In Eq. (18) the matrix, $\boldsymbol{W}^m(\mu,\mu')$, can be expanded in series of generalized spherical functions as (Garcia and Siewert, 1986)

$$\boldsymbol{W}^m(\mu,\mu') = \sum_{\ell=m}^{M} \boldsymbol{\Pi}_\ell^m(\mu)\boldsymbol{B}_\ell\boldsymbol{\Pi}_\ell^m(\mu'), \tag{22}$$

with

$$\boldsymbol{\Pi}_\ell^m(\mu) = \begin{bmatrix} Y_\ell^m & 0 & 0 & 0 \\ 0 & R_\ell^m & -T_\ell^m & 0 \\ 0 & -T_\ell^m & R_\ell^m & 0 \\ 0 & 0 & 0 & Y_\ell^m \end{bmatrix}, \tag{23}$$

where $Y_\ell^m$ is the normalized associated Legendre polynomials, $R_\ell^m$ and $T_\ell^m$ are defined in terms of the normalized spherical functions. The normalized Legendre polynomials are defined in terms of the associated Legendre polynomials $p_\ell^m(\mu)$ as (Dave, 1975)

$$Y_\ell^m(\mu) = \left[\frac{(\ell-m)!}{(\ell+m)!}\right]^{1/2} p_\ell^m(\mu) \qquad (24)$$

The upper limit of Eq. (22) defines the degree of anisotropic of the scattering matrix or the number of nonzero coefficients in its expansion in series of generalized spherical functions. The 4 by 4 matrices $\boldsymbol{B_\ell}$ are the coefficients of the expansion of the scattering matrix in series of the generalized spherical functions and are defined by the Greek constants $\{\gamma_\ell, \beta_\ell, \alpha_\ell, \zeta_\ell, \varepsilon_\ell, \delta_\ell\}$, so that

$$\boldsymbol{B_\ell} = \begin{bmatrix} \gamma_\ell & \beta_\ell & 0 & 0 \\ \beta_\ell & \alpha_\ell & 0 & 0 \\ 0 & 0 & \zeta_\ell & \varepsilon_\ell \\ 0 & 0 & -\varepsilon_\ell & \delta_\ell \end{bmatrix}. \qquad (25)$$

It should be noted that the properties of the generalized spherical functions, the recursion formulae for computing them and the relations for the computation of the Greek constants are given in detail in (Siewert, 1982; De Rooij, 1985).

## 4. Discrete spherical harmonics transform of the VRTE

Following the expansion of the phase matrix given in Eq. (18), the Stokes vector is expanded in Fourier series of the azimuthal angle as (Garcia and Siewert, 1986)

$$\boldsymbol{I}(\tau, \mu, \varphi - \varphi_0) = \sum_{m=0}^{\infty} \sum_{k=1}^{2} \Phi_k^m(\varphi - \varphi_0) \mathbf{I}_k^m(\tau, \mu). \qquad (26)$$

The coefficients of expansion define the vector radiance moment. Inserting Eq. (18) and (26) into Eq. (7) and (8) yields to the following VRTE for each Fourier component $m$

$$\mu \frac{\mathbf{I}_k^m(\tau, \mu)}{d\tau} + \mathbf{I}_k^m(\tau, \mu) = \frac{\omega(\tau)}{2} \int_{-1}^{1} \boldsymbol{W}^m(\mu, \mu') \mathbf{I}_k^m(\tau, \mu') \, d\mu' + \boldsymbol{Q}_k(\tau, \mu, \mu_0), \qquad (27)$$

where

$$\boldsymbol{Q}_k(\tau,\mu,\mu_0) = S_0(\tau) + \frac{\omega(\tau)}{4}\boldsymbol{W}^m(\mu,\mu_0)\boldsymbol{D}_k\boldsymbol{F}_0 e^{-\tau/\mu_0}. \tag{28}$$

Equation (27) can be solved independently for each $k = 1, 2$ and the subscript $k$ will be dropped down for the further calculations. It should be noted that Eq. (27) for each $m$ is identical to the equation of radiative transfer in a polarized inhomogeneous planar atmosphere with azimuthal symmetry. In the similar way as Eq. (22) the Fourier coefficients $\boldsymbol{I}^m(\tau,\mu)$ can be approximated as

$$\boldsymbol{I}^m(\tau,\mu) = \sum_{\ell=m}^{L+m} \boldsymbol{\Pi}_\ell^m(\mu)\,\boldsymbol{I}_\ell^m(\tau). \tag{29}$$

The upper limit of Eq. (29) is chosen according to Evans (1993). The odd number L refers to the order of the discrete spherical harmonics method.

In the DSHM hypothesis Eq. (22) and (29) are inserted in the VRTE of radiance moments, Eq. (27), and the resulting equation is solved for a set of polar directions. That is

$$\sum_{\ell=m}^{L+m}\mu\boldsymbol{\Pi}_\ell^m(\mu)\frac{d\boldsymbol{I}_\ell^m(\tau)}{d\tau} + \sum_{\ell=m}^{L+m}\boldsymbol{\Pi}_\ell^m(\mu)\boldsymbol{I}_\ell^m(\tau) = \omega(\tau)\sum_{\ell=m}^{L+m}\boldsymbol{\Pi}_\ell^m(\mu)\widetilde{\boldsymbol{B}_\ell}\boldsymbol{I}_\ell^m(\tau) = \boldsymbol{Q}(\tau,\mu,\mu_0) \tag{30}$$

In calculating the integral term in the right hand side of Eq. (27) the following orthogonal relation has been used (Siewert and McCormick, 1993)

$$\int_{-1}^{1}\boldsymbol{\Pi}_\ell^m(\mu)\boldsymbol{\Pi}_k^n(\mu)d\mu = \frac{2}{2\ell+1}\boldsymbol{\Delta}_\ell\delta_{k\ell}\delta_{mn}, \tag{31}$$

with

$$\boldsymbol{\Delta}_\ell = diag\{1,(1-\delta_{0\ell})(1-\delta_{1\ell}),(1-\delta_{0\ell})(1-\delta_{1\ell}),1\} \tag{32}$$

and

$$\widetilde{\boldsymbol{B}_\ell} = \boldsymbol{B}_\ell\boldsymbol{\Delta}_\ell/(2\ell+1). \tag{33}$$

The second main point of the DSHM transform of the VRTE consists to consider Eq. (30) for each polar direction $\mu_j$ as

$$\sum_{\ell=m}^{L+m} \mu_j Y_\ell^m(\mu_j) \frac{dI_\ell^m(\tau)}{d\tau} + \sum_{\ell=m}^{L+m} Y_\ell^m(\mu_j) I_\ell^m(\tau)$$

$$= \omega(\tau) \sum_{\ell=m}^{L+m} \Pi_\ell^m(\mu_j) \widetilde{B_\ell} I_\ell^m(\tau) = Q(\tau, \mu_j, \mu_0), \tag{34}$$

where $-J \leq j \leq J$, $J$ and integer. In order to have a system of $(L+1)$ coupled differential equations in $(L+1)$ unknowns in vector moments, the value of $J$ must be chosen in such a way that $2J = L+1$. For each Fourier component and for a given order of approximation Eq. (34) can be put in a matrix form as

$$[A]\frac{dI(\tau)}{ds} + [B]I(s) + \omega(s)[C]I(s) = Q(s), \tag{35}$$

where the following change of variable has been used $s = 2\tau/\tau_0 - 1$, $s \in [-1,1]$. The matrix $A$, $B$, $C$ are block matrices and $Q$ is a colonn block vector. For solving Eq. (35) with respect to the spatial variable, the boundary conditions Eq. (9) and (10) must also be transformed by the DSHM to read

$$\sum_{\ell=m}^{L+m} \Pi_\ell^m(\mu_j) I_\ell^m(-1) = 0, j = 1,2,\ldots J \ ; \ k = 1,2, \tag{36}$$

$$\sum_{\ell=m}^{L+m} [\Pi_\ell^m(\mu_j) - \beta_\ell^m] I_\ell^m(1) = \mu_0 \lambda_0 T I_0 e^{-\frac{\tau_0}{\mu_0}} \delta_{0,m} \ , j = 1,2,\ldots J \ ; \ k = 1 \tag{37}$$

and

$$\sum_{\ell=m}^{L} [\Pi_\ell^m(\mu_j) - \beta_\ell^m] I_\ell^m(1) = 0, j = 1,2,\ldots J; \ k = 2, \tag{38}$$

where

$$\beta_\ell^m = 2\lambda_0 T \int_0^1 \Pi_\ell^m(\mu') \mu' d\mu'. \tag{39}$$

## 5. Spatial discretization

To solve the spatial problem of the VRTE, Eq. (35), for a $m$ Fourier component the spectral Chebyshev method (STM) is adopted. The STM consists to expand the vector moments of radiance and its derivative in a truncated series of Chebyshev polynomials as (Shen et al., 2011)

$$\boldsymbol{I}(s) = \sum_{k=0}^{N_c} \Psi_k T_k(s) \tag{40}$$

and

$$\frac{\partial \boldsymbol{I}(s)}{\partial s} = \sum_{k=0}^{N_c} \Psi_k^{(1)} T_k(s), \tag{41}$$

where $N_c$ is the number of Chebyshev nodes defined by the relation $s_k = \cos\left(k\frac{\pi}{N_c}\right)$, $\Psi_k$ and $\Psi_k^{(1)}$ are the coefficients in Chebyshev expansion of the vector moments of radiance and its derivative, respectively. The coefficients, $\Psi_k^{(1)}$, are given in terms of the coefficients, $\Psi_k$, by the following relation (Shen et al., 2011)

$$\Psi_k^{(1)} = \frac{1}{1+\delta_{0k}} \sum_{\substack{p=k+1 \\ p+k \text{ odd}}}^{N_c} 2p\Psi_p. \tag{42}$$

The Chebyshev polynomials have the following properties $T_k(\pm 1) = (\pm 1)^k$ ) and are orthogonal so that (Shen et al., 2011)

$$\int_{-1}^{1} T_r(s) T_q(s) (1-s^2)^{-1/2} \, ds = \frac{\pi}{2}(1+\delta_{0r})\delta_{rq}. \tag{43}$$

By analogy to Eq. (40) the following expressions of $\omega(s)$ and $\boldsymbol{Q}(s)$ are written

$$\omega(s) = \sum_{q=0}^{N_c} \Omega_q T_q(s) \tag{44}$$

and

$$\boldsymbol{Q}(s) = \sum_{k=0}^{N_c} \boldsymbol{\Lambda}_k T_k(s). \tag{45}$$

Here, $\boldsymbol{\Lambda}_k$ and $\Omega_q$ are respectively the source term and the scattering albedo expansion coefficients in series of Chebyshev polynomials. When substituting Eqs. (40) to (42), (44) and (45) into Eq. (35) and applying the operator $\int_{-1}^{1} \cdots T_r(s)(1-s^2)^{-1/2} ds$ leads to

$$\frac{[\boldsymbol{A}^m]}{1+\delta_{0r}} \sum_{\substack{p=r+1 \\ p+r\, odd}}^{N_c} 2p\Psi_p + [\boldsymbol{B}^m]\Psi_r + \sum_{k=0}^{N_c} [\boldsymbol{C}^m]\gamma_{rk}\Psi_k = \boldsymbol{\Lambda}_r \ , r = 0 \ldots N_c, \tag{46}$$

where

$$\gamma_{rk} = \frac{2}{\pi(1+\delta_{0r})} \sum_{q=0}^{N_c} \Omega_q \int_{-1}^{1} T_q(s) T_k(s) T_r(s)(1-s^2)^{-1/2} ds. \tag{47}$$

In the same manner, the boundary conditions for the resolution of Eq. (46) read

$$\sum_{k=0}^{N_c} [\Gamma_+^m]\Psi_k T_k(-1) = \sum_{k=0}^{N_c} (-1)^k [\Gamma_+^m]\Psi_k = G_1^m \tag{48}$$

and

$$\sum_{k=0}^{N_c} [\Gamma_-^m]\Psi_k T_k(1) = \sum_{k=0}^{N_c} [\Gamma_-^m]\Psi_k = G_2^m \tag{49}$$

The solution of Eq. (46), (48) and (49) allows to determine the Chebyshev coefficients and therefore the radiance moments vector. Consequently, the total intensity and the Stokes vector characterizing the polarization state in the atmosphere can easily be deduced.

## 6. Numerical results

Two test problems are considered in this paper for the DSHM predictions. For the both problems, it is assumed that the atmosphere is illuminated at the top by the sunlight in the direction $\mu_0 = 0.2$, having the incident flux vector $\boldsymbol{I}_0 = [1,0,0,0]^T$. The matrix of reflectance is given by $T = diag\{1,0,0,0\}$. The inhomogeneous atmosphere is characterized by the single scattering albedo $\omega(\tau) = \omega_0 \exp(-a\tau)$ (Mishchenko, 1990), where $\omega_0 \in\, ]0,1]$ and $a \in [0,1]$ are the constants. Results are presented here for three values of the azimuthal angle, $\varphi = \{0, \pi/2, \pi\}$, which informs one on how the Stokes parameters vary over the plane. Also, four values of $a$ are considered, say $a = \{0, 0.01, 0.1, 1\}$. The order of the DSHM used in this calculation is $N = 35$ which corresponds to 36 steams. In addition, twenty nodes of Chebyshev were also used for the spatial discretization of the problems. The homogeneous atmosphere corresponds to the case $a = 0$. This later case is used in this work to show that the DSHM is well implemented and can be used for the radiative transfer predictions in polarized atmosphere.

The first problem considered is that of molecular scattering by the atmosphere. The scattering is conservative and the Rayleigh scattering law is given in table 1. The atmosphere of optical thickness $\tau_L = 0.5$ is bounded at the bottom by a non-reflecting surface. Figure 1 presents at the boundaries of the atmosphere the scattering intensity of the polarized radiation for different values of the azimuthal angle. Since the fourth Stokes parameter is zero for molecular scattering, calculation indicates that at azimuthal angles of 0° and 180° the third component vanishes. In this situation, the axes of the ellipse defining the plane of polarization are along the meridian plane. Figure 1 shows that the scattering polarized intensity is greatly affected by the value of the constant $a$. The lowest value of the radiance is observed for $a = 1$. As $a$ approaches zero this value increases, the corresponding curve is closed to the curve of $a = 0$ and the atmosphere behaves as a homogeneous slab medium. This aspect is fundamental to demonstrate that the DSHM presented in this work was well implemented. Moreover, a comparison between the DSHM predictions with the results obtained by Natraj et al. (2009) using a method based on the $X$ and $Y$ functions for homogeneous atmosphere also shows the validity of the method. Another aspect to verify the accuracy of the DSHM is the calculation of the relative difference in intensity which is less than 0.5% for $|\mu| \geq 0.1$ except for the directions close to zero where a maximum relative error of less than 3% is found. The last graph at the left of figure 1 informs one on how the radiative intensity is affected by the azimuthal angle.

The second example used in this paper is the scattering by polydispersed spherical particles. The atmosphere of optical thickness $\tau_L = 1$ is illuminated at the top by a solar radiation at $\lambda = 0.951 \mu m$. In addition, the atmosphere is bounded at the bottom by the ground of surface albedo $\lambda_0 = 0.1$. The size of the particles is described by a gamma distribution with the effective radius $r_{eff} = 0.2 \mu m$, the effective variance $v_{eff} = 0.07$ and the refractive index of $n = 1.44$ (Deirmendjian, 1969). The Mie theory is used to compute the Greeks constant for this problem and the results are listed in table 2. These coefficients agree to ten decimal places to those reported by Vestrucci and Siewert (1984). The constant scattering albedo is slightly modified to consider the absorption by the atmosphere. Thus the scattering albedo used for computation of the results is $\omega_0 = 0.99$. Figure 2 presents at the boundaries of the atmosphere the scattering intensity of the polarized radiation for different values of the azimuthal angle. In figure 3 the degree of polarization of the scattered light is also presented. As in the Rayleigh scattering, figure 2 shows that the scattered intensity is greatly affected by the value of the constant $a$. The lowest values of the intensity are obtained for $a = 1$, the value at which the degree of polarization shown in figure 3 is highest. As $a$ approaches zero the intensity increases, the degree of polarization decreases, the corresponding curve is closed to the curve of $a = 0$ and the atmosphere behaves as a homogeneous slab medium. This result is a first point indicating that the DSHM can be applicable for radiative transfer. Note that at the azimuthal angle $\varphi = 0$ and $\varphi = \pi$, the fourth Stokes component vanishes and the degree of polarization is reduced to the degree of linear polarization. Garcia and Siewert (1989) used the $F_N$ method for solving this problem for the homogeneous atmosphere ($a = 0$). The DSHM predictions are compared to their results and a good agreement is found. Once more, the relative difference is calculated to check how far the DSHM results are deviated from the $F_N$ results. The absolute value of the maximum deviation is less than 0.9% for $|\mu| \geq 0.1$ in the radiative intensity and less than 0.8% in the degree of polarization. For directions $\mu \to 0$ a maximum deviation less than 10% and 7% is observed respectively in radiative intensity and the degree of polarization. This result is not surprising since the SHM needs high order to converge for the directions close to zero.

**Conclusion**

The radiative transfer through a plane-parallel inhomogeneous polarized atmosphere is studied using a discrete spherical harmonics method. In the development of the method, the Stokes vector and the phase matrix are expanded in a finite series of the generalized spherical functions. Then, the resulting first order differential equations of vector moments are expressed in a set of discrete polar directions. The combination of the spherical harmonics and the discrete ordinates principles make the proposed method useful for radiative intensity and the degree of polarization predictions at any polar direction without using an interpolation function or linearization as it is generally encountered in discrete ordinates method. The resulting equations from the discrete spherical harmonics ordinates were put in a matrix form and the Chebyshev spectral method was adopted for its spatial discretization. Accuracy of the method was tested in the case of homogeneous atmosphere for both Rayleigh and Mie scattering. It was found that for a low order of the discrete spherical harmonics method, the emergent/transmitted radiative intensity and the degree of polarization predictions agree well with literature benchmark results except for polar directions close to zero where high order of angular discretization is required for accurate predictions.

**References**

Ackerman, T. P., Stokes, G. M., The atmospheric radiation measurement program. *Phys. Today*, **56**, 38 (2003).

Barlakas, V., Macke, A., and Wendisch, M., SPARTA-Solver for polarized atmospheric radiative transfer applications: Introduction and application to Saharan dust fields. *J. Quant. Spectrosc. Radiat. Transfer*, **178**, 77 (2016).

Benassi, M., Garcia, R. D. M., and Siewert, C. E., A generalized spherical harmonics solution basic to the scattering of the polarized light. *Journal of Applied Mathematics and Physics*, **36**, 70 (1985).

Bohren, C. F., and Clothiaux, E. E., Fundamentals of Atmospheric Radiation: An Introduction with 400 Problems. Wiley-VCH Verlag GmbH & Co. KGaA, Weinheim (2006).

Budak, V. P. Zheltov, V. S. Lubenchenko, A. V. Freidlin, K. S. Shagalov, O. V., A Fast and Accurate Synthetic Iteration-Based Algorithm for Numerical Simulation of Radiative Transfer in a Turbid Medium, *Atmospheric and Oceanic Optics*, **30** (1), pp. 70 (2017).

Budak, V.P., Korkin, S.V., The vectorial radiative transfer equation problem in the small angle modification of the spherical harmonics method with the determination of the solution smooth part, *Remote Sensing of the Atmosphere and Clouds*. Edited by Tsay, Si-Chee; Nakajima, Teruyuki; Singh, Ramesh P.; Sridharan, R.. Proc. of SPIE, Vol. 6408, 64081I (2006).

Collin, C., Pattanaik, S., LiKamWa, P., Bouatouch, K., Discrete ordinate method for polarized light transport solution and subsurface BRDF computation. Computers & Graphics, **45**, 17 (2014).

Cornet, C., Labonnote, L. C., Szczap, F., Three-dimensional polarized Monte Carlo atmospheric radiative transfer model (3DMCPOL): 3D effects on polarized visible reflectances of a cirrus cloud. *J. Quant. Spectrosc. Radiat. Transfer*, **111**, 174, (2010).

Dave, J., A Direct solution of the spherical harmonics approximation to the radiative transfer equation for an arbitrary solar elevation. Part I: Theory. *J. Atmos. Sci*, **32**, 790 (1975).

De Rooij, W. A., *Reflection and transmission of polarized light by planetary atmospheres*. PhD thesis, Vrije Universiteit te Amsterdam (1985).

Deirmendjian, D., *Electromagnetic scattering on spherical polydispersions*. American Elsevier Publishing Company, INC., New York (1969).

Dombrovsky, L.A. and Baillis, D., Thermal Radiation in Disperse Systems: An Engineering Approach, Begell House, New York (2010).

Evans, K. F., and Stephens, G. L., A new polarized atmospheric radiative transfer model. *J. Quant.Spectrosc.Radiat. Transfer*, **46** (5), 413 (1991).

Evans, K. F., Two-dimensional radiative transfer in cloudy atmospheres: The spherical harmonics spatial grid method. *J. Atmos. Sci*, **50** (18), 3111 (1993)

Garcia, R. D. M., and Siewert, C. E., A simplified implementation of the discrete-ordinates method for a class of problems in radiative transfer with polarization. *J. Quant. Spectrosc. Radiat. Transfer*, **112**, 2801 (2011).

Garcia, R. D.M., and Siewert, C. E., A generalized spherical harmonics solution for radiative transfer models that include polarization effects. *J. Quant. Spectrosc. Radiat. Transfer*, **36** (5), 401 (1986).

Garcia, R. D.M., and Siewert, C. E., The $F_N$ method for radiative transfer models that include polarization effects. *J. Quant. Spectrosc. Radiat. Transfer*, **41** (2), 117 (1989).

Hansen, J. E., and Travis, L. D., Light scattering in planetary atmospheres. *Space Science Reviews*, **16**, 527 (1974).

Hovenier, J. W., Multiple scattering of polarized light in planetary atmospheres. *Astron. Astrophys.*, **13**, 7 (1971).

Hovenier, J. W., Symmety relationships for scattering of polarized light in a slap of randomly oriented particles. *J. Atmos. Sci*, **26**, 488 (1969)

Korkin, S. V., Lyapustin, A. I., Vladimir V. Rozanov, V. V., APC: A new code for Atmospheric Polarization Computations. *J. Quant. Spectrosc. Radiat. Transfer*, **127**, 1 (2013).

Marshak, A., Davis, A., *3D radiative transfer in cloudy atmospheres*. Springer-Verlag, Berlin (2005)

Mishchenko, M. I., The fast invariant imbedding method for polarized light: computational aspects and numerical results for Rayleigh scattering. *J. Quant. Spectrosc. Radiat. Transfer*, **43** (3), 163 (1990).

Mishchenko, M. I., Travis, L. D., and Lacis, A. A., *Scattering, Absorption, and Emission of Light by Small Particles*. New York (2004)

Mishchenko M.I., Travis L.D., and Lacis, A.A., Multiple Scattering of Light by Particles: Radiative Transfer and Coherent Backscattering, Cambridge Univ. Press, Cambridge (UK) (2006)

Muñoz, A. G., and Mills, F. P., Pre-conditioned backward Monte Carlo solutions to radiative transport in planetary atmospheres. Fundamentals: Sampling of propagation directions in polarising media, *J. Astron. Astrophys,* **573**, 1 (2015).

Natraj, V., Li, K. F., and Yung; Y. L., Rayleigh scattering in planetary atmospheres: corrected tables through accurate computation of *X* and *Y* functions. *J. Astrophys*, **691**, 909 (2009)

Rozanov, V. V., and Kokhanovsky, A. A., The solution of the vector radiative transfer equation using the discrete ordinates technique: Selected applications. *Atmospheric Research*, **79**, 241 (2006).

Schulz, F. M. Stamnes, K., Weng, F., VDISORT: An improved and generalized discrete ordinate method for polarized (vector) radiative transfer, *J. Quant. Spectrosc. Radiat. Transfer* **61**(1), 105 (1999).

Shen, J., Tang, T., and Wang, L-L., *Spectral Methods Algorithms, Analysis and Applications*, Springer-Verlag Berlin Heidelberg (2011).

Siewert, C. E., and McCormick, N . J., A particular solution for polarization calculations in radiative transfer. *J. Quant. Spectrosc. Radiat. Transfer*, **50** (5), 513 (1993)

Siewert, C. E., On the phase matrix basic to the scattering of polarized light. *J. Astron. Astrophys.*, **109**, 195 (1982).

Spurr, R., LIDORT and VLIDORT: Linearized pseudo-spherical scalar and vector discrete ordinate radiative transfer models for use in remote sensing retrieval problems, In: Kokhanovsky A.A. (eds) Light Scattering Reviews 3. Springer Praxis Books. Springer, Berlin, Heidelberg, 229-275 (2008).

Stam, D. M., De Rooij, W. A., Cornet, G., and Hovenier, J. W., Integrating polarized light over a planetary disk applied to starlight reflected by extrasolar planets. *J. Astron. Astrophys.*, **452**, 669 (2006).

van de Hulst, H. G., *Light scattering by small particles*. Dover Publications, Inc., New York (1981).

Vestrucci, P., and Siewert, C. E., A Numerical evaluation of an analytical representation of the components in a Fourier decomposition of the phase matrix for the scattering of polarized light. *J. Quant. Spectrosc. Radiat. Transfer*, **31** (2), 177 (1984).

Wauben, W. M. F., and Hovenier, J. W., Polarized radiation of an atmosphere containing randomly-oriented spheroids. *J. Quant. Spectrosc. Radiat. Transfer*, **47** (6), 491 (1992).

Wauben, W. M. F., de Haan, J. F., and Hovenier, J. W., Low orders of scattering in a plane-parallel homogeneous atmosphere. *J. Astron. Astrophys*, **276**, 589 (1993).

Table 1: Rayleigh scattering law

| $\ell$ | $\gamma$ | $\beta$ | $\alpha$ | $\zeta$ | $\varepsilon$ | $\delta$ |
|---|---|---|---|---|---|---|
| 0 | 1 | 0 | 0 | 0 | 0 | 0 |
| 1 | 0 | 0 | 0 | 0 | 0 | 1.5 |
| 2 | 0.5 | 1.22474487 | 3 | 0 | 0 | 0 |

Table 2: Greeks constant

| $\ell$ | $\gamma$ | $\beta$ | $\alpha$ |
|---|---|---|---|
| 0 | 1.000000000000 | 0 | 0 |
| 1 | 1.455293181922 | 0 | 0 |
| 2 | 1.054026312815 | -0.755249151756 | 3.309122046421 |
| 3 | 0.397589937874 | -0.361993431909 | 0.963375827659 |
| 4 | 0.116593016009 | -0.115574881450 | 0.247412425589 |
| 5 | 0.023874769121 | -0.024981587107 | 0.045263695073 |
| 6 | 0.003950101559 | -0.004167534803 | 0.006889259618 |
| 7 | 0.000538878521 | -0.000573902106 | 0.000879817671 |
| 8 | 0.000063715321 | -0.000067787877 | 0.000098722959 |
| 9 | 0.000006668975 | -0.000007091825 | 0.000009901965 |
| 10 | 0.000000632596 | -0.000000670881 | 0.000000906787 |
| 11 | 0.000000055199 | -0.000000058381 | 0.000000076807 |
| 12 | 0.000000004483 | -0.000000004728 | 0.000000006084 |
| 13 | 0.000000000342 | -0.000000000360 | 0.000000000454 |
| $\ell$ | $\zeta$ | $\varepsilon$ | $\delta$ |
| 0 | 0 | 0 | 0.712063424566 |
| 1 | 0 | 0 | 1.760141193141 |
| 2 | 2.577320744341 | 0.042072687493 | 1.066824310779 |
| 3 | 0.757443760450 | 0.085067155506 | 0.396511038967 |
| 4 | 0.163817766550 | 0.015431841988 | 0.095764123767 |
| 5 | 0.027831477813 | 0.003153487404 | 0.017650880723 |
| 6 | 0.003889722669 | 0.000401029552 | 0.002615486256 |
| 7 | 0.000464262578 | 0.000046013589 | 0.000327130182 |
| 8 | 0.000049020814 | 0.000004286900 | 0.000035829893 |
| 9 | 0.000004664063 | 0.000000361426 | 0.000003513577 |
| 10 | 0.000000407359 | 0.000000027190 | 0.000000314654 |
| 11 | 0.000000033041 | 0.000000001874 | 0.000000026068 |
| 12 | 0.000000002517 | 0.000000000118 | 0.000000002022 |
| 13 | 0.000000000181 | 0.000000000007 | 0.000000000148 |

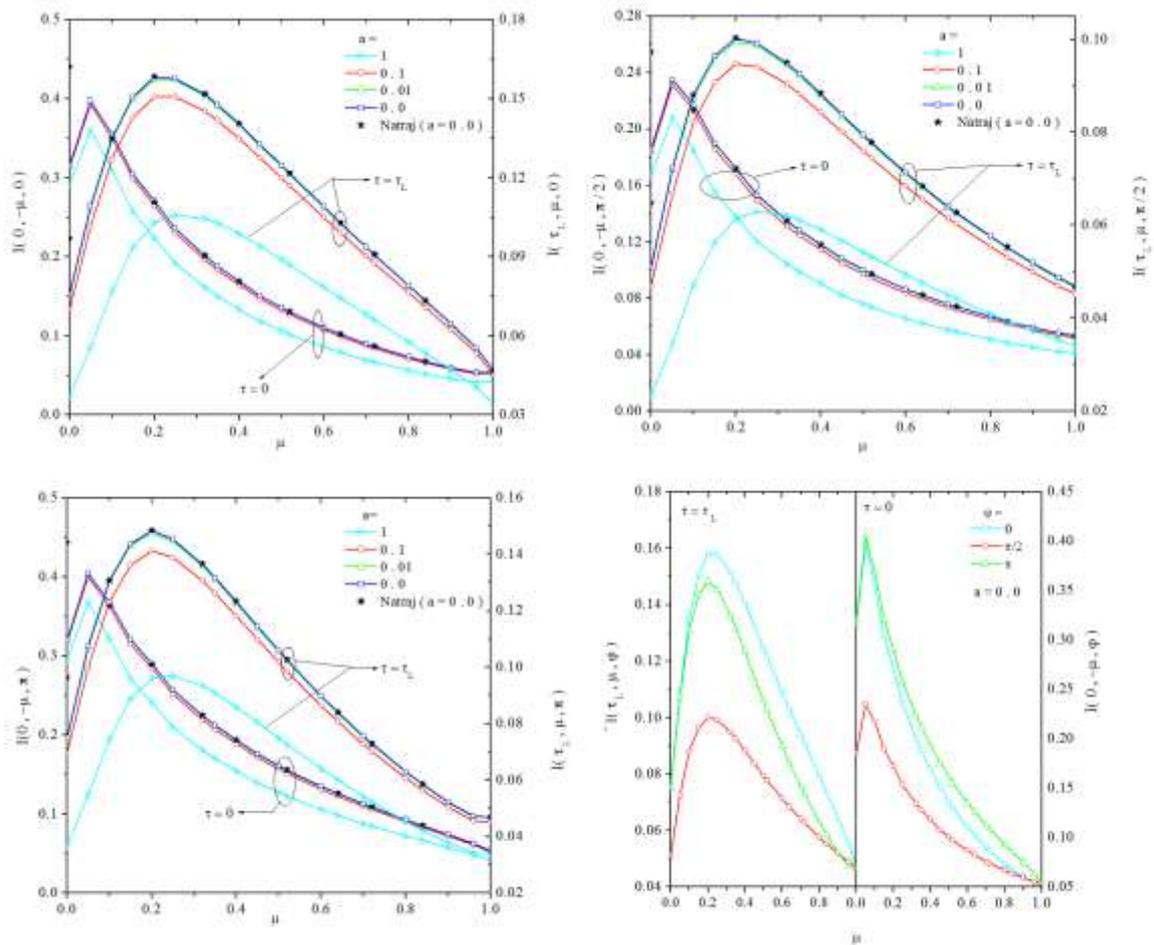

Figure 1: Emergent and transmitted intensity of scattering by molecules.

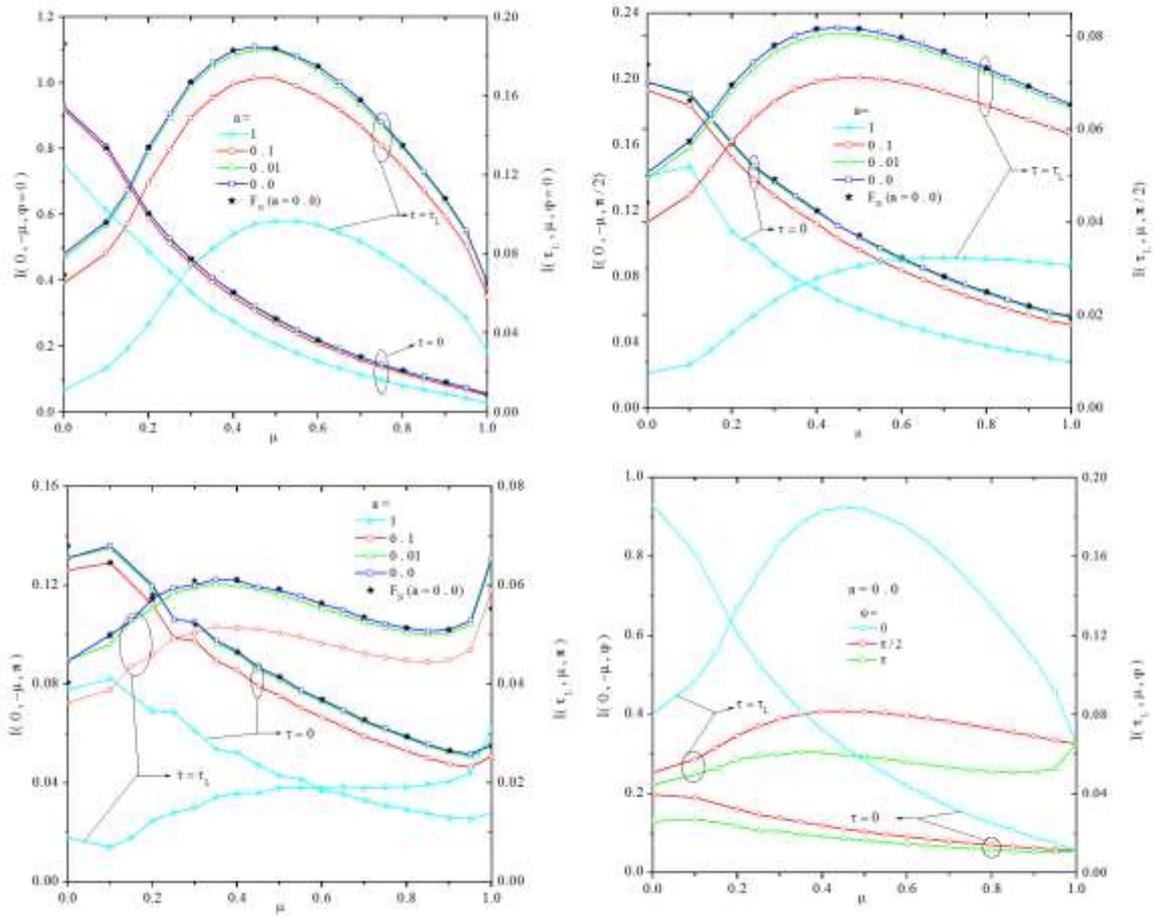

Figure 2: Emergent and transmitted intensity of scattering by polydispersed sphere.

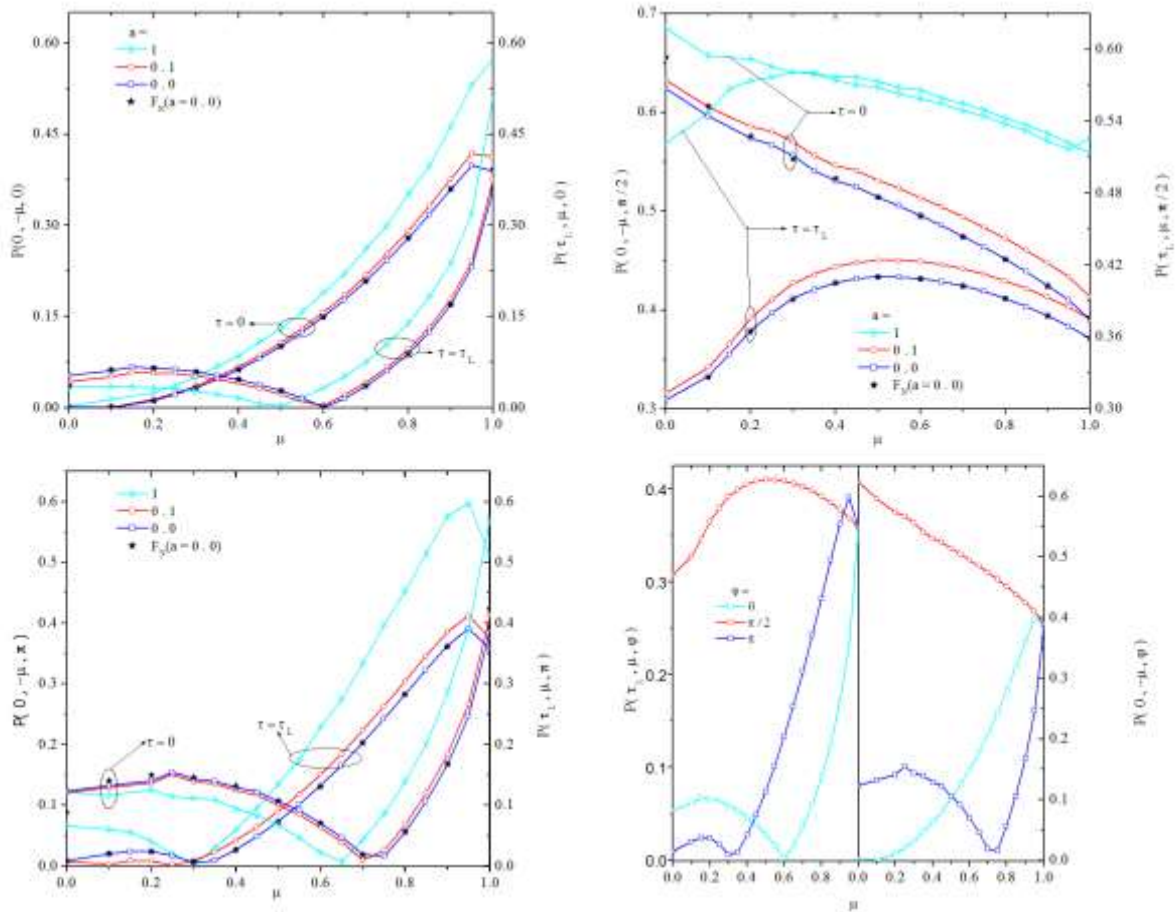

Figure 3: Degree of polarization of the scattering by polydispersed sphere.